\documentclass[twocolumn,showpacs,preprintnumbers,amsmath,amssymb]{revtex4}

\usepackage{graphicx}% Include figure files

\begin{document}

\title{Boson-boson effective nonrelativistic potential for higher-derivative

electromagnetic theories in D dimensions}

\author{Antonio Accioly}

\email{accioly@ift.unesp.br}

\author{Marco Dias }

 \affiliation{Instituto de F\'{\i}sica
Te\'{o}rica, Universidade Estadual Paulista, Rua Pamplona 145,
01405-000 S\~ao Paulo, SP, Brazil}

\date{\today}

\begin{abstract}

 The problem of computing the effective nonrelativistic potential
$U_{D}$ for the interaction
 of charged scalar bosons within the
 context of D-dimensional electromagnetism with a cutoff, is
reduced to quadratures. It is shown that $U_3$
 cannot bind a pair of identical charged scalar bosons;
 nevertheless, numerical calculations indicate that boson-boson
 bound states do exist in the framework of three-dimensional
  higher-derivative electromagnetism augmented by a topological Chern-Simons term.

\end{abstract}


 \maketitle

 \section{introduction}

 We consider in this Brief Report the problem of determining the effective
charged-scalar-boson---charged-scalar-boson low energy potential
 $U_D$
 arising from D-dimensional electromagnetism with a cutoff $a$.
 The Lagrangian concerning this theory can be written as

\begin{eqnarray}
{\cal L} = -\frac{1}{4} F_{\mu\nu} F^{\mu\nu} + \frac{a^2}{2}
\partial_\nu F^{\mu\nu} \partial^\lambda F_{\mu\lambda},
\end{eqnarray}

\noindent where  $F_{\mu\nu} \equiv \partial_\nu A_\mu -
\partial_\mu A_\nu$ is the usual electromagnetic tensor field.
Lagrangian (1) is gauge and Lorentz invariant; in addition, it
leads to local field equations which are linear in the field
quantities. At distances much larger than the cutoff, the fields
described by Eq. (1) become essentially equivalent to the Maxwell
fields.

It worth mentioning that Lagrangian (1) was proposed a long time
ago, in 3+1 dimensions, by Podolsky and Schwed \cite{1}, in a
rather different context. The main reason for investigating this
theory is that recently it was shown that in its framework the
electromagnetic mass of a point charge occurs in the equation of
motion in a form consistent with special relativity; moreover, the
exact equation of motion does not exhibit runaway solutions or
non-causal behavior, when the cutoff is larger than half of the
classical radius of the electron \cite{2,3}. Massive fermions, in
turn, have their helicity flipped on account of their interaction
with an electromagnetic field described by Podolsky's generalized
electrodynamics, while massless ones seem to be unaffected by the
electromagnetic field as far as their helicity is concerned
\cite{4}. Now, as is well known, if one starts from a
renormalizable theory and calculates the effective action one
obtains typically several operators of dimension 6 (or higher) and
not just Podolsky's term \cite{5}. This raises the question: Would
these latter results change significantly if one takes all these
terms into account? Since the amount of spin-flip is determined by
the vector nature of the electromagnetic field but not by its
spatial distribution, the aforementioned results will not change
at all. What about the finiteness of the electromagnetic mass?
Since the evaluation of the classical self-force acting on the
point charge depends on the cutoff $a$ \cite{2,3}, one is led to
conjecture that the finiteness of the electromagnetic mass is a
very specific property of Podolsky's electrodynamics and not a
general feature of higher-derivative theories. It is interesting
that the same route that leads to Maxwell's electrodynamics leads
also to Podolsky's electrodynamics provided we start from
Podolsky's electrostatic force instead of the usual Coulomb's law
\cite{6,7}. Unlike Maxwell's  electrodynamics, Podolsky's
generalized electrodynamics implies a finite value for the energy
of a point charge in the whole space \cite{8,9}.

We are motivated by two quite similar developments: In the first,
we investigate whether  $U_3$  can form ``Cooper pairs".

Our second topic is related to three-dimensional
Podolsky-Chern-Simons theory. Based on the interesting discussions
from Jackiw \cite{10} about the  consistency of the
nonrelativistic limit of certain relativistically invariant
quantum field theories, it can be shown that the Chern-Simons term
alone is unable to form boson-boson bound states \cite{11}.
Nonetheless, numerical calculations indicate that the Podolsky
term provides an stabilizing mechanism allowing for the existence
of ``Cooper pairs".

We use natural units throughout; our signature is
$(+,-,-,\cdot\cdot\cdot,-)$.

\section{effective charged-scalar-boson---charged-scalar-boson low energy potential}

 \indent We begin by describing a method for computing the
 effective nonrelativistic potential for the interaction of two
 charged spinless bosons of equal masses via a ``Podolskian
 photon" exchange. The prescription is based on the marriage of
 nonrelativistic quantum mechanics and quantum field theory in the
 nonrelativistic limit. An algorithm for calculating the
 propagator is then presented. The recipe is used afterward  to
 get the propagator for higher-derivative electromagnetism in the
 Lorentz gauge. Finally, we reduce the problem of computing the
 effective nonrelativistic potential to quadratures.

\subsection{The method}

\indent Nonrelativistic quantum mechanics tells us that in the
first Born approximation the cross section for the scattering of
two indistinguishable massive particles, in the center-of-mass
frame (CoM), is given by $\frac{d\sigma}{d\Omega} = \left |
\frac{m}{4\pi} \int e^{-i {\bf p'} \cdot \; {\bf r}} V( r)
e^{i{\bf p} \cdot\; {\bf r}} d^{D-1}\;{\bf{r}} \right |^2,$ where
${\bf p}\; ({\bf p'})$ is the initial (final) momentum of one of
the particles in the CoM. In terms of the transfer momentum, ${\bf
k \equiv p' - p}$, it reads

\begin{eqnarray}
\frac{d\sigma}{d\Omega} = \left | \frac{m}{4\pi} \int V( r)
e^{i{\bf k} \cdot\; {\bf r}} d^{D-1}\;  {\bf{r}} \right |^2.
\end{eqnarray}

\indent On the other hand, from quantum field theory we know that
the cross section, in the CoM, for the scattering of two identical
massive scalars bosons by an electromagnetic field, can be written
a $\frac{d\sigma}{d\Omega} = \left | \frac{1}{16\pi E}\; {\cal
M}\right |^2,$ where $E$ is the initial energy of one of the
bosons and ${\cal M}$ is the Feynman amplitude for the process at
hand, which in the nonrelativistic limit  $(N.R.)$ reduces to

\begin{eqnarray}
\frac{d\sigma}{d\Omega} = \left | \frac{1}{16\pi m}\; {\cal
M}_{N.R.} \right |^2.
\end{eqnarray}

\indent From Eqs. (2) and (3) we come to the conclusion that the
expression that enables us to compute the D-dimensional effective
nonrelativistic potential has the form

\begin {eqnarray}
V( r) = \frac{1}{4m^2} \frac{1}{(2\pi)^{D-1}} \int d^{D-1}\; {\bf
k}\; {\cal M}_{N.R.}\; e^{-i {\bf k} \cdot \; {\bf r}},
\end{eqnarray}

\noindent which clearly shows how the potential from quantum
mechanics and the Feynman amplitude obtained via quantum field
theory are related to each other.

\indent Now, in the Lorentz gauge Podolsky's scalar QED is
described by the Lagrangian

\begin{eqnarray}
{\cal L} &=& -\frac{1}{4} F_{\mu\nu} F^{\mu\nu} + \frac{a^2}{2}
\partial_\nu F^{\mu\nu} \partial^\alpha
F_{\mu\alpha}-\frac{1}{2\lambda} (\partial_\nu A^\nu)^2
\nonumber\\ &&+ (D_\mu \phi)^* D^\mu \phi - m^2 \phi^* \phi,
\end{eqnarray}

\noindent where $D_\mu \equiv \partial_\mu + iQA_\mu$. Therefore,
the interaction Lagrangian to order $Q$ for the process $S + S
\longrightarrow S + S$, where $S$ denotes a spinless boson of mass
$m$ and charge $Q$, is ${\cal L}_{int} = iqQ A^\mu \left( \phi
\partial_\mu \phi^* - \phi^*
\partial_\mu \phi \right).$ The Feynman rule for the elementary vertex
is shown in Fig. 1. Accordingly, the Feynman amplitude for the
interaction of two charged spinless bosons of equal mass via a
``Podolskian photon" exchange (see Fig. 2) is

\begin{center}

\begin{figure}[h]

\begin{tabular}{cc}

\includegraphics[scale=.1]{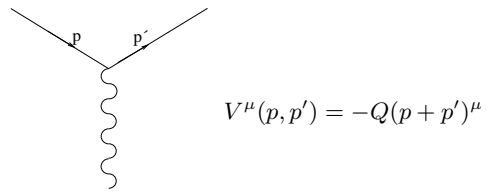}&

\begin{tabular}{l}

 $V^\mu(p,p')=-Q(p+p')^\mu$\\ \\ \\ \\ \\ \\

\end{tabular}

\end{tabular}

\caption{The relevant vertex for boson-boson interaction.}

\end{figure}

\end{center}

\begin{center}

\begin{figure}[b]

\includegraphics[scale=.1]{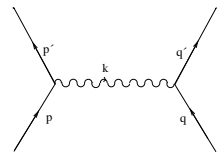}

\caption{One-Podolskian photon-exchange contibution to the
scattering of two identical massive charged bosons.}

\end{figure}

\end{center}

\begin{eqnarray}
{\cal M} =V^{\mu} (p,p') D_{\mu\nu} (k) V^\nu (q,q'),
\end{eqnarray}

\noindent where $D_{\mu\nu} (k)$ designates the ``Podolskian
photon" propagator.

\subsection{The propagator}

\indent We propose now an algorithm for computing the propagator
for electromagnetic theories with higher-derivatives, based on the
usual transverse and longitudinal vector projector operators,
namely $\theta_{\mu\nu} = \eta_{\mu\nu} - \frac{\partial_\mu
\partial_\nu}{\Box},\; \omega_{\mu\nu} = \frac{\partial_\mu
\partial_\nu}{\Box},$ which satisfy the relations $\theta_{\mu\rho} {\theta^\rho}_\nu =
\theta_{\mu\nu},\;\omega_{\mu\rho} {\omega^\rho}_\nu =
\omega_{\mu\nu},\;\theta_{\mu\rho} {\omega^\rho}_\nu = 0,$ where
$\eta_{\mu\nu}$ is the Minkowski metric. The set of operators $\{
\theta,\omega\}$ is a complete set of projector operators for
rank-one tensors. Indeed, they are idempotent, mutually orthogonal
and satisfy the completness relation $[\theta + \omega]_{\mu\nu} =
\eta_{\mu\nu} \equiv I_{\mu\nu}.$

  Let ${\bar{\cal L}}$ be the Lagrangian for electromagnetism with
  higher-derivatives. Since ${\bar{\cal L}}$ is a gauge -invariant
  Lagrangian, we add to it a gauge-fixing Lagrangian ${\cal
 L}_{gf}$, which implies that ${\cal L} \equiv {\bar{\cal L}} +
 {\cal L}_{gf}$ can be written as ${\cal L} = \frac{1}{2} A^{\mu} O_{\mu\nu} A^\nu.$
  Expanding $O$ in the basis $\{ \theta,\omega\}$, yields $O = x_1 \theta + x_2 \omega.$
  Accordingly, $O^{-1} = y_1 \theta + y_2 \omega,$ where
$O^{-1}$ is the propagator and $y_1$ and $y_2$ are parameters to
be determined. Now, taking into account that $O O^{-1} = I,$
 we promptly obtain

 $$O^{-1} = \frac{1}{x_1} \theta + \frac{1}{x_2} \omega,$$

\noindent where we are supposing that both $x_1$ and $x_2$ are
nonvanishing. Note that the procedure we have just outlined is
quite straightforward: on the one hand it reduces the work of
calculating the propagator to a trivial algebraic exercise; on the
other hand  it  great simplifies calculations involving the
contraction of conserved currents $(\partial_\nu J^\nu = 0)$ with
the propagator since in this case the alluded contraction simply
gives

\begin{eqnarray}
{O^{-1}}_{\mu\nu} J^\mu = \frac{J_\nu}{x_1}.
\end{eqnarray}

From the above we find that the propagator for Podolsky's
electrodynamics in the Lorentz gauge assumes the form

\begin{eqnarray}
D_{\mu\nu} (k) = \frac{M^2}{k^2 (k^2 - M^2)}\; \theta_{\mu\nu}
-\frac{\lambda}{k^2}\; \omega_{\mu\nu},
\end{eqnarray}

\noindent where $M^2 \equiv \frac{1}{a^2}.$

\subsection{The potential}

From Eqs. (6), (7) and (8), we get immediately ${\cal M} =
\frac{M^2 Q^2 (2p -k)\cdot(2q + k)}{k^2 (k^2-M^2)},$ which implies

\begin{eqnarray}
{\cal M}_{N.R.} = \frac{4m^2M^2Q^2}{{\bf k}^2 ({\bf k}^2 + M^2)}.
\end{eqnarray}

Inserting Eq. (9) into Eq. (4), we obtain

\begin{eqnarray}
V(r) &=& {\int_0}^{\;\infty} f(|{\bf k}|) |{\bf k}|^{n-1}\; d|{\bf
k}|\; {\int_0}^{\; 2\pi} d\theta_1 \nonumber \\ &&\times
{\int_0}^{\; \pi}\sin \theta_2 \; d\theta_2 \;{\int_0}^{\; \pi}
{\sin ^2} \theta_3 \; d\theta_3 \nonumber  \\ &&... {\int_0}^{\;
\pi} e^{-i |{\bf k}| r \cos \theta_{n-1}} \sin^{n-2}
\theta_{n-1}\; d\theta_{n-1}, \nonumber
\end{eqnarray}

 \noindent where $2 <n \equiv D-1$ and $ f(|{\bf k}|) \equiv
 \frac{Q^2}{(2 \pi)^n} \left( \frac{1}{\;{\bf k}^2} - \frac{1}{\;{\bf
 k}^2 - M^2} \right).$ Now, taking into account that

\begin{eqnarray}
{\int_0}^{\; \pi} \sin^{m} \theta \; d\theta = \frac{\sqrt{\pi}\;
\Gamma \left( \frac {m - 1}{2} \right)}{\Gamma \left( \frac{ m +
2}{2} \right)}, \nonumber
\end{eqnarray}

\begin{eqnarray}
 {\int_0}^{\; \pi} e^{-i |{\bf k}| r
\cos \theta_{n-1}} \sin^{n-2} \theta_{n-1}\; d\theta_{n-1}&=&
\frac{ 2^{\frac {n-2}{2}}\; \Gamma \left(\frac{1}{2}
\right)}{\left( |{\bf k}| r
 \right)^{\frac{n-2}{2}}} \nonumber \\ && \times
 \Gamma \left( \frac{n-1}{2} \right) \nonumber \\ && \times J_{\frac{n-2}{2}} \left(|{\bf k}| r
 \right), \nonumber
\end{eqnarray}

\noindent where $J$ denotes the Bessel function, we arrive at the
following expression for the potential

\begin{widetext}

\begin{equation}
U_{D}(r) = \frac{Q}{(2\pi)^{\frac{D-1}{2}}\; r^{\frac{D-3}{2}}}
{\int_0}^{\;\infty} \left( \frac{1}{\;{\bf k}^2} - \frac{1}{\;{\bf
k}^2 -M^2} \right)|{\bf k}|^{\frac{D-1}{2}}\; J_{\frac{D-3}{2}}
\left(|{\bf k}| r \right) \; d|{\bf k}|,
\end{equation}

\end{widetext}

\noindent where $U_{D} (r) \equiv  \frac{V(r)}{Q}$ and $D>3.$

On the other hand , it is trivial to show that $U_3$ can be
evaluated from the expression

\begin{eqnarray}
U_3(r) = \frac{Q}{2\pi} {\int_0}^{\;\infty} \left( \frac{1}{\;{\bf
k}^2} - \frac{1}{\;{\bf k}^2 -M^2} \right)|{\bf k}| J_{0}\left(
|{\bf k}| r \right) \; d|{\bf k}|,\nonumber
\end{eqnarray}

 \noindent which allows us to conclude that Eq. (10) can also be
 applied to the case $D=3$. Hence, the problem of computing the effective nonrelativistic
potential for $D>2$ was reduced to quadratures.

If Eq. (10) is correct , it must reproduce the Podolskian
potential in 3+1 dimensions. Performing the computation for $D=4$,
we get

\begin{eqnarray}
U_4 (r) = \frac{Q}{4 \pi} \frac{1 - e^{ - \frac{r}{a}}}{r},
\nonumber
\end{eqnarray}

\noindent which is just the same result as that obtained in
Podolsky's electromagnetic theory [1].

\section{Planar quadratic electromagnetism}

For $D=3$, Eq. (10) yields

\begin{eqnarray}
U_3(r) = -\frac{Q}{2 \pi} \left[ \ln {\frac{r}{r_0}} + K_0(Mr)
\right],
\end{eqnarray}

 \noindent where $r_0$ is an infrared regulator and $K$ is the
 modified Bessel function.

We discuss now the existence of boson-boson bound states in the
context of planar quadratic electromagnetism. The corresponding
time-independent Schr\"odinger equation can be written as

\begin{eqnarray}
{\cal{H}}_l {\cal{R}}_{nl} &=& -\frac{1}{m} \left( \frac{d^2}{d
r^2} {\cal{R}}_{nl} + \frac{1}{r} \frac{d}{d r} {\cal{R}}_{nl}
\right) + V^{eff}_l {\cal{R}}_{nl} \nonumber \\ &=& E_{nl} {\cal
R}_{nl}, \nonumber
\end{eqnarray}

\begin{eqnarray}
V^{eff}_l &\equiv& \frac{l^2}{m r^2} + Q U_3(r) \nonumber \\ &=&
\frac{l^2}{m r^2} - \frac{Q^2}{2\pi} \left[\ln{\frac{r}{r_0}} +
K_0(Mr) \right], \nonumber
\end{eqnarray}

\noindent where ${\cal R}_{nl}$ is the $n$th normalizable
eigenfunction of the radial Hamiltonian ${\cal H}_l$ whose
corresponding eigenvalue is $E_{nl}$ and $V^{eff}_l$ is the $l$th
partial wave effective potential. On the other hand,

\begin{eqnarray}
\frac{d}{d r} V^{eff}_l = -\frac{2l^2}{m} \frac{1}{r^3} -
\frac{Q^2}{2 \pi} \frac{1}{r} + \frac{Q^2 M}{2
\pi}K_1(Mr),\nonumber
\end{eqnarray}

\noindent which allows us to conclude that $\frac{d}{d r}
V^{eff}_l < 0$ in the interval $ 0 < r < \infty$, implying that
$V^{eff}_l$ is strictly decreasing in this interval. Consequently,
in the framework of planar quadratic electromagnetism, no bound
state concerning the two charged scalar bosons system exists.

\section{podolsky-chern-simons planar electromagnetism}
Since boson-boson bound states do not show up in Podolsky planar
electromagnetism, we investigate here whether the effective
boson-boson low energy potential related to Podolsky-Chern-Simons
(PCS) planar theory can bind a pair of identical charged scalar
bosons. The Lagrangian for PCS scalar QED, in the Lorentz gauge,
can be written as

\begin{eqnarray}
{\cal L} &=& -\frac{1}{4} F_{\mu\nu} F^{\mu\nu} + \frac{a^2}{2}
\partial_\nu F^{\mu\nu} \partial^\alpha
F_{\mu\alpha}-\frac{1}{2\lambda} (\partial_\nu A^\nu)^2
\nonumber\\ &&+ (D_\mu \phi)^* D^\mu \phi - m^2 \phi^* \phi
+\frac{s}{2} \varepsilon_{\mu\nu\rho} A^\mu \partial ^\nu A^\rho,
\end{eqnarray}

 \noindent where $s>0$ is the topological mass.

In the basis $\{\theta,\omega,S\}$, where $S_{\mu\nu} \equiv
\varepsilon_{\mu\rho\nu} \partial^\rho$, the propagator assumes
the form

\begin{eqnarray}
O^{-1} = \frac{(a^2 k^4-k^2) \theta}{\left( a^2k^4-k^2 \right)^2 -
s^2 k^2} -\frac{\lambda \omega}{k^2} - \frac{s S}{\left(
a^2k^4-k^2 \right)^2- s^2k^2} . \nonumber
\end{eqnarray}

Now, in the nonrelativistic limit the Feynman amplitude for the
process shown in Fig. 2 reduces to

\begin{eqnarray}
{\cal M}_{NR} =  \left[ \frac{(a^2{\bf k}^4 + {\bf k}^2
)4Q^2m^2}{(a^2{\bf k}^4 + {\bf k}^2 )^2 + s^2{\bf k}^2} +
\frac{8ismQ^2 \; {\bf k} \wedge {\bf P} }{(a^2{\bf k}^4 + {\bf
k}^2 )^2 + s^2{\bf k}^2} \right], \nonumber
\end{eqnarray}

\noindent where ${\bf P} \equiv \frac{1}{2} ({\bf p} -{\bf q})$ is
the relative momentum of the incoming charged scalar bosons in the
CoM. On the other hand, is trivial to show  that if $a < \frac{2
\sqrt 3}{9s}$ the equation

\begin{eqnarray}
x^3 + \frac{2x^2}{a^2}  + \frac{x}{a^4} + \frac{s^2}{a^4} = 0,
\end{eqnarray}

\noindent where $ x \equiv {\bf k}^2 $, has three distinct
negative real roots. In this case  ${\cal M}_{NR}$ can be
rewritten as

\begin{eqnarray}
{\cal M}_{NR}&=&  \frac{8ismQ^2\;{\bf k} \wedge {\bf P}}{a^4}
\left[ \sum_{j=1}^{3} \frac{ B_j}{{\bf k}^2 -x_j} + \frac{a^4}{s^2
{\bf k}^2} \right] \nonumber \\ &&+ \frac{4Q^2 m^2}{a^4}
\sum_{j=1}^{3} \frac{A_j}{{\bf k}^2 - x_j}, \nonumber
\end{eqnarray}

\noindent where $x_1, x_2$ and $x_3$ are the roots of Eq. (13) and
$A_1 \equiv \frac{1 +a^2 x_1}{(x_1-x_2)(x_1-x_3)}$\;,\;$A_2 \equiv
\frac{1 +a^2 x_2}{(x_2-x_1)(x_2-x_3)}$\;, \;$A_3 \equiv \frac{1
+a^2 x_3}{(x_3-x_1)(x_3-x_2)}$\;,\;$B_1 \equiv
\frac{-(1+a^2x_1)^2}{s^2(x_1-x_2)(x_ 1-x_3)}$\; ,\;$B_2 \equiv
\frac{-(1+a^2x_2)^2}{s^2(x_2-x_1)(x_2-x_3)}$\;,\;$B_3 \equiv
\frac{-(1+a^2x_3)^2}{s^2(x_3-x_1)(x_3-x_2)}$.

\vskip .5cm

It follows that the effective nonrelativistic potential can be
calculated from the expression

\begin{eqnarray}
U_3(r) &=& \frac{i s Q}{\pi m a^4} \left[ \frac{a^4}{s^2}
\lim_{\sigma \rightarrow 0} \int_{0}^{\infty} \frac{({\bf k}
\wedge {\bf P}) J_0({|\bf k|}r) |{\bf k}|\; d|{\bf k}|}{{\bf k}^2
+ \sigma^2} \right. \nonumber \\
 &&+ \left. \sum_{j} \int_{0}^{\infty} \frac{({\bf k} \wedge {\bf P}) B_j J_0(|{\bf k}|r)
|{\bf k}|\; d|{\bf k}|}{{\bf k}^2 - x_j}  \right] \nonumber \\ &&+
\frac{Q}{2 \pi a^4} \sum_{j} \int_{0}^{\infty} \frac{A_j}{{\bf
k}^2 - x_j} J_0(|{\bf k}|r) |{\bf k}| d|{\bf k}|. \nonumber
\end{eqnarray}

Performing the computations, we obtain

\begin{widetext}

\begin{eqnarray}
U_3(r) = - \frac{s Q}{\pi m a^4}\left[ \frac{ a^4}{s^2}
\frac{1}{r^2} + \frac{1}{r} \sum_{j} B_j {\sqrt{|x_j|}}
\;K_1({\sqrt{|x_j|}}\; r) \right] {\bf L } + \frac{Q}{2 \pi a^4}
\left[ \sum_{j} A_j K_0({\sqrt{|x_j|}}\; r) \right],
\end{eqnarray}

\end{widetext}

\noindent where ${\bf L} \equiv {\bf r} \wedge {\bf P}$ is the
orbital angular momentum.

Using Jackiw's arguments \cite {10}, one can show that the
topological term alone is unable to bind the charged scalars
bosons \cite {11}.

We return now to the problem of probing whether ``Cooper pairs"
exist in the framework of PCS scalar QED. In this case the radial
Schr\"odinger equation is

\begin{equation}
\left[\frac{d^2}{dr^2}+\frac{1}{r} \frac{d}{dr} \right]{\cal R}
_{nl}+m\left[E_{nl}-V^{eff}_l\right]{\cal R}_{nl}=0,
\end{equation}

\noindent where

\begin{widetext}

\begin{eqnarray}
V^{eff}_l(r) = - \frac{s Q^2}{\pi m a^4}\left[ \frac{ a^4}{s^2}
\frac{1}{r^2} + \frac{1}{r} \sum_{j} B_j {\sqrt{|x_j|}}
\;K_1({\sqrt{|x_j|}}\; r) \right]l  + \frac{Q^2}{2 \pi a^4} \left[
\sum_{j} A_j K_0({\sqrt{|x_j|}}\; r) \right] + \frac{l^2}{m r^2}.
\nonumber
\end{eqnarray}

\end{widetext}

\noindent Employing the dimensionless parameters $ y \equiv sr $\;
,\;$ \alpha \equiv \frac{Q^2}{\pi s}$\; ,\; $b_j \equiv
\frac{s^2}{a^4}  B_j$\; ,\; $X_j \equiv \frac{|x_j|}{s}$ \; ,\;
$\beta \equiv \frac{m}{s}$ \; ,\; $a_j \equiv \frac{A_j}{a^4}$ and
$ {\tilde E}_{nl} \equiv \frac{m E_{nl}}{s^2}$, we can rewrite Eq.
(15) as

\begin{eqnarray}
\left[\frac{d^2}{dy^2}+\frac{1}{y} \frac{d}{d y} \right]{\cal R}
_{nl}+\left[{\tilde E}_{nl}-{\tilde V}^{eff}_l\right]{\cal
R}_{nl}=0,
\end{eqnarray}

 \noindent with

\begin{eqnarray}
{\tilde V}^{eff}_l &\equiv& - \frac{l(\alpha -l)}{y^2} +
\frac{\alpha\beta}{2} \sum_{j} a_j K_0(X_j\; y)  \nonumber \\ &&-
\frac{\alpha l}{y} \sum_{j} b_j X_j K_1(X_j \;y).\nonumber
\end{eqnarray}

\noindent Note that ${\tilde V}^{eff}_l$ behaves as
$\frac{l^2}{y^2}$ at the origin and as $\frac{l(l-\alpha)}{y^2}$
asymptotically. On the other hand, the derivative of this
potential with respect to $y$ is given by

\begin{eqnarray}
\frac{d}{d y}{\tilde V}^{eff}_l &=& \frac{2l(\alpha -l)}{y^3} +
\alpha \sum_{j}\left[ \frac{2l}{y^2} b_j - \frac{\beta a_j}{2}
\right ] X_j K_1(X_j \; y) \nonumber \\ &&+ \frac{\alpha l}{y}
\sum_{j} b_j X^2_j K_0(X_j  \;y). \nonumber
\end{eqnarray}

\noindent In order to find out whether or not boson-boson bound
states could be formed, we shall analyze how $\frac{d}{d y}{\tilde
V}^{eff}_l$ behaves for small values of the cutoff $a$. Indeed,
only if $a \lll 1$ will the well recognized properties of $QED_3$
be preserved. In this limit, we get

\begin{eqnarray}
\frac{d}{d y}{\tilde V}^{eff}_l \sim \frac{2l(\alpha -l)}{y^3} -
\left[ \frac{2 \alpha l}{y^2} + \frac{\alpha \beta}{2} \right]
K_1(y) - \frac{\alpha l}{y}K_0(y).\nonumber
\end{eqnarray}

\noindent We assume from now on $a \lll 1$ and $l> 0$, without any
loss of generality. It is trivial to see that if $l > \alpha$, the
potential is strictly decreasing, which precludes the existence of
bound states. The remaining possibility is $l < \alpha$. In this
interval ${\tilde V}^{eff}_l$ approaches $ + \infty$ at the origin
and $0^{-}$ for $y \rightarrow  + \infty$, which is indicative of
a local minimum. Therefore, the existence of ``Cooper pairs" is
subordinated to the conditions $a \lll 1$ and $0<l< \alpha$.

Of course, it is impossible to solve Eq. (16) analytically;
however, it  can be solved numerically. To do that, we rewrite
beforehand the radial function as $ {\cal R}_{nl} \equiv
\frac{u_{nl}}{\sqrt{y}}$. As a consequence, Eq. (16) takes the
form

\begin{eqnarray}
\left[ \frac{d^2}{d y^2} + \frac{1}{4 y^2} \right] u_{nl} + \left[
{\tilde E}_{nl} - {\tilde V}^{eff}_l \right] u_{nl} = 0.
\end{eqnarray}

\noindent Using the Numerov algorithm \cite {12}, we solved Eq.
(17) numerically for several values of the parameters $\alpha
\;,\; \beta$ and $l$, keeping the cutoff $a$ fixed. The latter was
chosen equal to $\frac{2}{3} r_e = 9.52033 \times 10^{-3}
MeV^{-1}$, where $r_e$ is the fourdimensional classical radius of
the electron. It is worth mentioning that the anomalous factor of
$\frac{4}{3}$ in the inertia related to the Abraham-Lorentz model
for the electron does not show up if if $a > \frac{1}{2} r_e$
[2,3].

In Fig. 3 we present our numerical results for the potential and
the corresponding radial eigenfuntions concerning the first three
bound states in the specific case of $l= 4$. The associated
energies are

$ E_{14} = - 6.37501 \times 10^{-7} MeV \;,\; E_{24} = -1.2536
\times 10^{-7} MeV \;,\; E_{34} = - 5.22441 \times 10^{-9} MeV.$

\noindent The graphics shown in Fig. 3 exhibit, in a sense, the
generic features of the potential and of the radial
eigenfunctions, although they have been composed using particular
values of the parameters $\alpha \;,\; \beta$ and $a$. A detailed
study of the modifications of the effective potential induced by
radioactive corrections, as well as the corresponding alterations
to the eigenvalue structure, will be published elsewhere \cite
{11}

To conclude, we remark that  ``Cooper pairs" exist in the context
of PCS scalar QED if $a \lll 1$ and $0< l < \alpha$.

\begin{center}

\begin{figure}[tb]

\includegraphics[scale=.7]{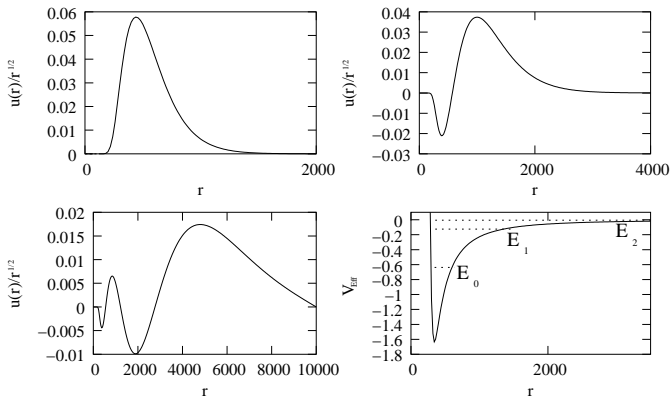}

\caption{$V^{eff}_4$ with the lowest three allowed energies and
the corresponding energy eigenfunctions. Here $\left[ V^{eff}_4
\right]=eV$, $[r]=MeV^{-1}$, $\alpha =8$, $\beta =2000$ and
$a=0.00952\,MeV^{-1}.$ }

\end{figure}

\end{center}

\begin{acknowledgments}

 A.A. thanks CNPq-Brazil and M.D. thanks CAPES-Brazil for financial support.

\end{acknowledgments}

\end{document}